\documentclass{ims9x6}

\usepackage{cite}

\newcommand*{\D}{\mathrm{d}}                      
\providecommand*{\I}{\mathrm{i}}                  
\providecommand*{\bra}[1]{{\left\langle#1\right|}}             
\providecommand*{\ket}[1]{{\left|#1\right\rangle}}             
\newcommand*{\tr}{\mathrm{tr}}			  
\newcommand*{\ds}{\displaystyle}                   
\newcommand*{\adj}{^{\dagger}}
\newcommand*{\Eq}[2][Eq.~]{#1(\ref{eq:EHHPT-#2})}
\newcommand*{\Eqs}[2]{Eqs.~(\ref{eq:EHHPT-#1})--(\ref{eq:EHHPT-#2})}
\newcommand*{\magn}[1]{\mathopen{\boldsymbol{|}}#1\mathclose{\boldsymbol{|}}}
\newcommand*{\Min}{\mathop{\mathrm{Min}}}
\newcommand*{\Max}{\mathop{\mathrm{Max}}}
\newcommand*{\Ext}{\mathop{\mathrm{Ext}}}
\newcommand*{\pder}[2]{\frac{\partial{#1}}{\partial{#2}}}
\newcommand*{\phadj}{^{\phantom{\dagger}}}
\newcommand*{\php}{^{\phantom{\prime}}}
\newcommand*{\pheq}{\mathrel{\phantom{=}}}
\newcommand*{\scaledmath}[2]{\scalebox{#1}{$\ds{#2}$}} 
\newcommand*{\sds}[2][\depth]{\raisebox{0pt}[\height][#1]{%
    \scaledmath{0.8}{#2}}}
\newcommand*{\Exp}[1]{\mathrm{e}^{\mbox{\small$\ds{#1}$}}}   

\DeclareMathAlphabet{\vecfont}{OT1}{cmr}{bx}{it}
\renewcommand{\vec}[1]{\vecfont{#1}}
\newcommand*{\grad}{\boldsymbol{\nabla}}

\usepackage{color}

\makeatletter
\renewcommand{\ps@plain}{%
  \renewcommand{\@oddhead}{\hfil\footnotesize%
    \raisebox{30pt}[0pt][0pt]{\parbox{300pt}{\centering%
      A contribution to the Proceedings of the\\{}%
      Workshop on Density Functionals for Many-Particle Systems\\{}
      2--27 September 2019, Singapore}}\hfil}%
  \renewcommand{\@evenhead}{\@oddhead}%
  \renewcommand{\@oddfoot}{\hfil\footnotesize%
        \raisebox{-8pt}[0pt][0pt]{\thepage}\hfil}%
  \renewcommand{\@evenfoot}{\@oddfoot}%
}
\makeatother
\renewcommand{\trimmarks}{\relax}

\begin{document}

\chapter{\uppercase{Energy~functionals of single-particle~densities: 
    A~unified view}}

\markboth%
{Berge~Englert, Alex~Hue, Jonah Huang, Miko\l{}aj Paraniak, and Martin~Trappe}%
{Energy functionals of single-particle  densities: A unified view}
\author{Berthold-Georg~Englert,$^{a,b,c}$
        Jun Hao Hue,$^{a,d}$ Zi Chao Huang,$^b$\\ %
        Miko\l{}aj M. Paraniak,$^a$ and Martin-Isbj\"orn~Trappe$^{a}$}
        
\address{$^a$Centre for Quantum Technologies, %
             National University of Singapore, Singapore\\%
         $^b$Department of Physics, %
             National University of Singapore, Singapore\\%
         \makebox[0pt][c]{$^c$MajuLab, CNRS-UCA-SU-NUS-NTU International %
                          Joint Research Unit, Singapore}\\
         $^d$Graduate School for Integrative Sciences \& Engineering, \\%
	     National University of Singapore, Singapore%
             \\[0.5ex]%
         englert@u.nus.edu, %
         junhao.hue@u.nus.edu, huangjonahzichao@gmail.com, %
          mikolajp@protonmail.ch, martin.trappe@quantumlah.org}

\begin{abstract}
Density functional theory is usually formulated in terms of the density in
configuration space.
Functionals of the momentum-space density have also been studied, and yet
other densities could be considered.
We offer a unified view from a second-quantized perspective and introduce a
version of density functional theory that treats all single-particle
contributions to the energy exactly.
An appendix deals with semiclassical eigenvalues.
\end{abstract}

\section{Introduction}
The very rich literature on density functional theory (DFT) consists almost
exclusively of articles on functionals of the configuration-space density and
their many applications; see Refs.~\cite{HSchapter,OBchapter} and the
references therein.
There are, however, experimentally accessible properties of many-electron
systems that require knowledge of the density in momentum space for their
computation --- the prime example are Compton profiles, the observed
wavelength distribution of photons scattered off electrons with a range of
initial velocities \cite{Compton:23}.
This triggered interest in an alternative version of DFT in terms of the
momentum-space density which produced a handful of publications~%
\cite{Henderson:81,Buchwald+Englert:89,Englert:92,Cinal+Englert:92,%
  Cinal+Englert:93,vConta+Siedentop:15}.
Perhaps it is worth considering functionals of yet other densities?

We present a unified view based on the second-quantized formalism for many
interacting identical particles.
The modes to which the creation and annihilation operators refer determine the
natural choice of the single-particle density, and we obtain the functionals
of this density by a constrained-search strategy of 
the Levy--Lieb kind \cite{Levy:79,Lieb:83}.
We recover the familiar functionals of the configuration-space and the
momentum-space densities and show how one can construct functionals of a third
kind where all single-particle contributions to the energy have an exactly
known functional while, as always, the functional for the contribution
of the pair interaction requires systematic approximations.

An appendix offers remarks on semiclassical approximations.
In particular, the familiar expressions for approximate WKB energies emerge
without reference to wave functions.

\section{Constrained search}
We consider systems of many identical particles with pair interactions, for
which the many-particle Hamilton operator $H_{\text{mp}}$ is the sum of
single-particle contributions and pair contributions, 
\begin{equation}\label{eq:EHHPT-B1}
  H_{\text{mp}}=H_{\text{single}}+H_{\text{pair}}\,,
\end{equation}
with
\begin{equation}\label{eq:EHHPT-B2}
  H_{\text{single}}=\sum_{a,a'}\psi(a)\adj\bra{a}H_{\text{1p}}\ket{a'}\psi(a')
\end{equation}
and
\begin{equation}\label{eq:EHHPT-B3}
  H_{\text{pair}}=\frac12\sum_{a,a',}\sum_{b,b'}\psi(a)\adj\psi(b)\adj
  \bra{a,b}H_{\text{int}}\ket{a',b'}\psi(b')\psi(a')\,.
\end{equation}
Here, $\psi(a)$ and $\psi(a)\adj$ are the annihilation and creation operators
for the mode labeled by $a$, $\ket{a}$ is the single-particle
ket for the $a$th mode,
${\bra{a}=\ket{a}\adj}$ is the adjoint bra, and
${\bra{a,b}\adj=\ket{a,b}=\ket{a}\otimes\ket{b}}$ is a two-particle
tensor-product ket; see, for example, Chapter 10 in
Ref.~\cite{Schwinger:QMbook}. 
The modes refer to a complete orthonormal set in the single-particle space,
\begin{equation}\label{eq:EHHPT-B3'}
  \langle a|b\rangle=\delta(a,b)\,,\qquad\sum_a\ket{a}\bra{a}=1\,,
\end{equation}
where
\begin{equation}
  \delta(a,b)={\left\{
      \begin{array}{c@{\ \text{if}\ }l}
        1&a=b\\0&a\neq b
      \end{array}\right\}}=\delta_{a,b}
\end{equation}
is the Kronecker delta symbol.\footnote{\label{fn:EHHPT-1}%
  The mode label could be continuous
  rather than discrete, or a combination of a continuous and a discrete
  label (position and spin, say), and then integrals replace the sums in
  \Eq[Eqs.~]{B2} and \Eq[]{B3}, and Dirac's delta function replaces the
  Kronecker delta symbol.
  We leave these matters implicit until we need to be explicit about them.} 
While \Eqs{B1}{B3} apply to systems of fermions or bosons, with
the respective algebraic properties of the $\psi(a)$s and their adjoints,
we focus on spin-$\frac12$ fermions here and, therefore, have the
anticommutators 
\begin{align}\label{eq:EHHPT-B4}
  \psi(a)\psi(b)+\psi(b)\psi(a)&=0\,,\nonumber\\
  \psi(a)\adj\psi(b)\adj+\psi(b)\adj\psi(a)\adj&=0\,,\nonumber\\
  \psi(a)\psi(b)\adj+\psi(b)\adj\psi(a)&=\delta(a,b)\,.
\end{align}

For the purpose of this paper, the generic form of the single-particle
energy
\begin{equation}\label{eq:EHHPT-B6}
  H_{\text{1p}}(\vec{R},\vec{P})
  =\frac{1}{2m}\vec{P}^2+V_{\text{trap}}(\vec{R})
\end{equation}
is the sum of the kinetic energy and the potential energy associated with the
conservative trapping forces;
$\vec{P}$ and $\vec{R}$ are the momentum and position vector operators of
a particle with mass $m$.
Further, we only consider pair interaction energies that result from
conservative line-of-sight forces,
\begin{equation}\label{eq:EHHPT-B7}
  H_{\text{int}}=V_{\text{int}}\bigl(\magn{\vec{R}_1-\vec{R}_2}\bigr)\,,
\end{equation}
where $\vec{R}_1$ and $\vec{R}_2$ are the position operators of any two
particles.
These restrictions can be lifted as the need arises.
In particular, one can account for spin-dependent contributions, such as those
of magnetic dipole-dipole interaction \cite{Goral+2:01,Fang+Englert:11}, or
one can replace $H_{\text{single}}$ of \Eq{B2} with $H_{\text{1p}}$ of \Eq{B6} by
Dirac's expression for relativistic fermions \cite{RMDchapter}.
Of course, one can also generalize to systems with particles of two or more
kinds. 

We choose the second-quantized version of the Hamilton operator in \Eqs{B1}{B3}
because it is more flexible than the first-quantized version with
\begin{equation}\label{eq:EHHPT-B8}
  H_{\text{single}}=\sum_{k=1}^N\biggl(
  \frac{1}{2m}\vec{P}_k^2+V_{\text{trap}}\bigl(\vec{R}_k\bigr)\biggr)
\end{equation}
and
\begin{equation}\label{eq:EHHPT-B9}
  H_{\text{pair}}=\frac12\mathop{\sum_{j,k=1}^N}_{(j\neq k)}
  V_{\text{int}}\bigl(\magn{\vec{R}_j-\vec{R}_k}\bigr)\,.
\end{equation}
In \Eq[Eqs.~]{B8} and \Eq[]{B9}, we have exactly $N$ particles;
for \Eq[Eqs.~]{B2} and \Eq[]{B3} this corresponds to only considering
many-particle states from one eigenvalue sector of the number operator
\begin{equation}
  \mathcal{N}=\sum_a\psi(a)\adj\psi(a)\,,
\end{equation}
that is
\begin{equation}
  \parbox[b]{0.52\textwidth}{\raggedright%
    $\mathcal{N}\ket{\ }=\ket{\ }N$
    for all permissible many-particle kets $\ket{\ }$.}
\end{equation}
The permissible kets are superpositions of the basic
$N$-particle kets in the Fock space,
\begin{equation}
  \psi(a_1)\adj\psi(a_2)\adj\cdots\psi(a_N)\adj\ket{\text{vac}}\,,
\end{equation}
where $N$ creation operators act on the vacuum ket $\ket{\text{vac}}$, which
describes the situation of no particles at all:
${\psi(a)\ket{\text{vac}}=0}$ for all modes, ${\mathcal{N}\ket{\text{vac}}=0}$.

We exploit the flexibility of the second-quantized formulation by minimizing
the energy under the constraint of a prechosen expectation value
of~$\mathcal{N}$,
\begin{equation}\label{eq:EHHPT-B13}
  \parbox[b]{0.52\textwidth}{\raggedright%
    $\tr\bigl(\mathcal{N}\rho\bigr)=N$
    for all permissible many- particle
    statistical operators $\rho\,,$}
\end{equation}
where $N$ is now any positive number, integer or noninteger.
As always, the requirements ${\rho\geq0}$ and ${\tr(\rho)=1}$ identify the set
of statistical operators.
Other than that, the $\rho$s are linear combinations of the basic building
blocks,
\begin{equation}\label{eq:EHHPT-B14a}
  \psi(a_1)\adj\psi(a_2)\adj\cdots\psi(a_K)\adj
  \ket{\text{vac}}\bra{\text{vac}}
  \psi(b_1)\psi(b_2)\cdots\psi(b_L)\,,
\end{equation}
with $K$ creation operators acting on the vacuum ket and $L$
annihilation operators acting on the vacuum bra
${\bra{\text{vac}}=\ket{\text{vac}}\adj}$.
We have
\begin{equation}
  \label{eq:EHHPT-B14b}
  \ket{\text{vac}}\bra{\text{vac}}
  =\prod_a\psi(a)\psi(a)\adj
\end{equation}
in view of the commutation relations in \Eq{B4}.
Only terms in $\rho$ with ${K=L}$ contribute to the expectation value of
$\mathcal{N}$ in \Eq{B13} and that of $H_{\text{mp}}$ in
\begin{equation}\label{eq:EHHPT-B15}
  E_{\text{gs}}(N)=\Min_{\rho}\bigl\{\tr(H_{\text{mp}}\rho)\bigr\}\,,
\end{equation}
where $H_{\text{mp}}$ is the Hamilton operator of \Eqs{B1}{B3}, and only
$\rho$s that obey the constraint in \Eq{B13} participate in the competition.%
\footnote{
  Yes, there are situations in which $\tr(H\rho)$ has an infimum but no
  minimum; for example, when there is no trapping potential and no interaction,
  so that we only have kinetic energy. We are not interested in these cases
  and shall assume that  $\tr(H\rho)$ has a minimum.
  If you feel uncomfortable with that, just read ``infimum'' for each
  occurrence of ``minimum.''}
As indicated, the ground-state energy $E_{\text{gs}}$ is a function of $N$; in
fact, it also depends on the particle mass, on the parameters that specify
$V_{\text{trap}}$ in \Eq{B6} and $V_{\text{int}}$ in \Eq{B7}, and on Planck's
constant that appears in the Heisenberg--Born commutation relation of the
components of $\vec{P}$ and $\vec{R}$,
\begin{equation}
  \I\bigl[\vec{a}\cdot\vec{P},\vec{b}\cdot\vec{R}\bigr]
  =\hbar\vec{a}\cdot\vec{b}\,,
\end{equation}
for any two numerical vectors $\vec{a}$ and $\vec{b}$.

The Levy--Lieb \cite{Levy:79,Lieb:83} constrained-search strategy finds
the minimum in \Eq{B15} in two steps.
First, we choose occupation numbers $n_a$ and restrict $\rho$ by the set of
constraints
\begin{equation}
  \label{eq:EHHPT-B16}
  \tr\bigl(\psi(a)\adj\psi(a)\rho\bigr)=n_a\,;
\end{equation}
then we consider all sets of $n_a$s such that
\begin{equation}
  \label{eq:EHHPT-B17}
  \sum_an_a=N\,.
\end{equation}
Clearly, this enforces the constraint \Eq[]{B13}.
The first step yields the energy functional
\begin{equation}
  \label{eq:EHHPT-B18}
  E[n]=\Min_{\rho\leadsto n}\bigl\{\tr(H_{\text{mp}}\rho)\bigr\}\,,
\end{equation}
where $n$, the \emph{single-particle density}, stands for the list of
occupation numbers and ${\rho\leadsto n}$ (``$\rho$ leads to $n$'')
symbolizes the constraints in \Eq{B16}; it is common practice
to call $E[n]$ the \emph{density functional} rather than the ``energy functional
of the density.''
In the second step, we have
\begin{equation}
  \label{eq:EHHPT-B19}
   E_{\text{gs}}(N)=\Min_{n\leadsto N}\bigl\{E[n]\bigr\}\,,
\end{equation}
where ${n\leadsto N}$ is the constraint in \Eq{B17}.
This two-step approach is useful if $E[n]$ lends itself to systematic
approximations --- the central challenge of DFT.

We are free to choose the modes, specified by the single-particle kets
$\ket{a}$ and their adjoint bras $\bra{a}$, at our convenience.
The occupation numbers $n_a$, and thus the density $n$, refer to this choice
and, therefore, the functional $E[n]$ depends on the choice as well.
As illustrated by the three particular choices in Secs.~\ref{sec:EHHPT-3},
\ref{sec:EHHPT-4}, \ref{sec:EHHPT-6} and the explicit Thomas--Fermi
functionals in Sec.~\ref{sec:EHHPT-5}, the structure of $E[n]$ depends on the
choice of modes very strongly. 
While we leave the mode dependence implicit and do not indicate it in the
notation, we must remember that the density $n$ and all density functionals
are context specific.

If we denote the minimizer in \Eq{B18} by $\rho[n]$,%
\footnote{When $N$ is integer, the density functional $\rho[n]$ is composed of
  terms with ${K=L=N}$ in \Eq{B13}; when $N$ is noninteger,
  ${N_1<N<N_2=N_1+1}$, $\rho[n]$ is composed of terms with ${K=L=N_1}$ or
  ${K=L=N_2}$; 
  see Refs.~\cite{Perdew+3:82,Perdew+3:00,Yang+2:00,Baerends:20} and
  references therein.}
then
\begin{align}\label{eq:EHHPT-B18'}
  E[n]=\tr\bigl(H_{\text{mp}}\rho[n]\bigr)
  &=\tr\bigl(H_{\text{single}}\rho[n]\bigr)
   +\tr\bigl(H_{\text{pair}}\rho[n]\bigr)\nonumber\\
  &=E_{\text{single}}[n]+E_{\text{pair}}[n]\,,
\end{align}
where $E_{\text{single}}[n]$ and $E_{\text{pair}}[n]$ are defined jointly, not
individually.
Both functionals change when we modify $V_{\text{trap}}(\vec{r})$ in
$H_{\text{single}}$ while not modifying $V_{\text{int}}(\magn{\vec{r}})$ in
$H_{\text{pair}}$, or modify $V_{\text{int}}(\magn{\vec{r}})$ in
$H_{\text{pair}}$ while not modifying $V_{\text{trap}}(\vec{r})$ in
$H_{\text{single}}$.
Upon introducing the reduced single-particle statistical operator
\begin{equation}\label{eq:EHHPT-B99a}
  n^{(1)}=\sum_{a,a'}\ket{a'}\,\tr\Bigl(\psi(a')\rho[n]
  \psi(a)\adj\Bigr)\bra{a}
  \quad\text{with}\quad\bra{a}n^{(1)}\ket{a}=n_a
\end{equation}
and the reduced two-particle statistical operator
\begin{equation}\label{eq:EHHPT-B99b}
  n^{(2)}=\frac12\sum_{a,a'}\sum_{b,b'}\ket{a',b'}\,
  \tr\Bigl(\psi(b')\psi(a')\rho[n]
  \psi(a)\adj\psi(b)\adj\Bigr)\bra{a,b}\,,
\end{equation}
which are functionals of the single-particle density $n$,
we have
\begin{equation}\label{eq:EHHPT-B99c}
  E_{\text{single}}[n]=\tr\Bigl(H_{\text{1p}}n^{(1)}\Bigr)
  \quad\text{and}\quad
  E_{\text{pair}}[n]=\tr\Bigl(H_{\text{int}}n^{(2)}\Bigr)
\end{equation}
for the density functionals of the single-particle energy and the pair
energy.
There are many-particle traces in \Eq[Eqs.~]{B18'}--\Eq[]{B99b}, and a
single-particle trace as well as a two-particle trace in \Eq{B99c}.

We incorporate the constraint of \Eq{B17} into the density functional with the
aid of a Lagrange multiplier $\mu$, the chemical potential,
\begin{equation}
  \label{eq:EHHPT-B20}
  E[n,\mu]=E[n]+\mu N-\mu\sum_an_a\,,
\end{equation}
and then the ground-state energy is the stationary value of $E[n,\mu]$,
\begin{equation}
  \label{eq:EHHPT-B21}
  E_{\text{gs}}(N)=\Ext_{n\leadsto N,\mu}\bigl\{E[n,\mu]\bigr\}\,.
\end{equation}
While this extremum can be a minimum, usually it is a saddle point.
In any case, the ground-state values of the density and the chemical
potential, $n_{\text{gs}}$ and $\mu_{\text{gs}}$, obey
\begin{equation}
  \pder{}{n_a}E[n]=\mu\,,\qquad N=\sum_an_a\,,
\end{equation}
so that
\begin{equation}
   E_{\text{gs}}(N)=E[n_{\text{gs}}]\,.
\end{equation}

\section{Configuration-space functionals}\label{sec:EHHPT-3}
In traditional DFT with its extensive literature \cite{HSchapter,OBchapter}
the emphasis is on the three-dimensional configuration space.%
\footnote{While there are one- and two-dimensional variants, also for the
  momentum-space functionals of Sec.~\ref{sec:EHHPT-4}, we elaborate on the
  three-dimensional case only.}
This corresponds to
\begin{equation}
  \psi(a)\to\psi_{\sigma}(\vec{r})\,,
\end{equation}
which annihilates a particle with spin label $\sigma$ at position $\vec{r}$.
The symbolic summation over the mode label $a$ is realized by summation over
$\sigma$ and integration over $\vec{r}$, as exemplified by the number operator
\begin{equation}\label{eq:EHHPT-C2}
  \mathcal{N}=\sum_{\sigma}\int(\D\vec{r})\,
  \psi_{\sigma}(\vec{r})\adj\psi_{\sigma}(\vec{r})\,,
\end{equation}
where $\sigma$ has two values for the spin-$\frac12$ fermions under
consideration and $(\D\vec{r})$ is the spatial volume element.
The third line in \Eq{B4} now reads
\begin{equation}\label{eq:EHHPT-C3}
  \psi_{\sigma}(\vec{r})\psi_{\sigma'}(\vec{r}')\adj
  +\psi_{\sigma'}(\vec{r}')\adj\psi_{\sigma}(\vec{r})
  =\delta_{\sigma,\sigma'}\delta(\vec{r}-\vec{r}')\,,
\end{equation}
as anticipated in footnote~\ref{fn:EHHPT-1}.

With $H_{\text{1p}}$ in \Eq{B6} and  $H_{\text{int}}$ in \Eq{B7}, we have
\begin{align}
  \bra{a}H_{\text{1p}}\ket{a'}
  &\to \bra{\vec{r},\sigma}\biggl(
    \frac{1}{2m}\vec{P}^2+V_{\text{trap}}(\vec{R})\biggr)
    \ket{\vec{r}',\sigma'}
    \nonumber\\
  &=\delta_{\sigma,\sigma'}
    \biggl(-\frac{\hbar^2}{2m}\grad^2+V_{\text{trap}}(\vec{r})\biggr)
    \delta(\vec{r}-\vec{r}')
\end{align}
and
\begin{align}
  \bra{a,b}H_{\text{int}}\ket{a',b'}
  &\to\bra{\vec{r}_1\php,\sigma_1\php;\vec{r}_2\php,\sigma_2\php}
    V_{\text{int}}\bigl(\magn{\vec{R}_1-\vec{R}_2}\bigr)
    \ket{\vec{r}_1',\sigma_1';\vec{r}_2',\sigma_2'}\nonumber\\
  &=\delta_{\sigma_1\php,\sigma_1'}\delta_{\sigma_2\php,\sigma_2'}
    \delta(\vec{r}_1\php-\vec{r}_1')\delta(\vec{r}_2\php-\vec{r}_2')
    V_{\text{int}}\bigl(\magn{\vec{r}_1-\vec{r}_2}\bigr)\,,
\end{align}
which yield
\begin{equation}\label{eq:EHHPT-C6}
  H_{\text{single}}=\sum_{\sigma}\int(\D\vec{r})\,\psi_{\sigma}(\vec{r})\adj
  \biggl(-\frac{\hbar^2}{2m}\grad^2+V_{\text{trap}}(\vec{r})\biggr)
  \psi_{\sigma}(\vec{r})
\end{equation}
and
\begin{equation}\label{eq:EHHPT-C7}
  H_{\text{pair}}=\frac12\sum_{\sigma,\sigma'}
  \int(\D\vec{r})(\D\vec{r}')\,
  \psi_{\sigma}(\vec{r})\adj\psi_{\sigma'}(\vec{r}')\adj
   V_{\text{int}}\bigl(\magn{\vec{r}-\vec{r}'}\bigr)
  \psi_{\sigma'}(\vec{r}')\psi_{\sigma}(\vec{r})  \,.
\end{equation}
The analog of \Eq{B16} is the spatial single-particle density
\begin{equation}\label{eq:EHHPT-C8}
  n(\vec{r})=\tr\biggl(\sum_{\sigma}
  \psi_{\sigma}(\vec{r})\adj\psi_{\sigma}(\vec{r})\rho\biggr)\,,
\end{equation}
where the two spin components are added.
It is also possible --- and, strictly speaking, more in line with \Eq{B16} ---
to use both spin components
\begin{equation}
  n_{\sigma}(\vec{r})=\tr\biggl(
  \psi_{\sigma}(\vec{r})\adj\psi_{\sigma}(\vec{r})\rho\biggr)
\end{equation}
and deal with the corresponding ``spin-density functionals;''
see Ref.~\cite{LZGchapter} and the references therein.
We consider only the usual spin-summed density of \Eq{C8} and note that
\begin{equation}\label{eq:EHHPT-C10}
  E_{\text{single}}[n]=E_{\text{kin}}[n]
  +\int(\D\vec{r})\,V_{\text{trap}}(\vec{r})n(\vec{r})
\end{equation}
has an exactly known functional for the potential energy of the trapping
forces plus a functional for the kinetic energy that is jointly defined with
the pair-energy functional,
\begin{equation}\label{eq:EHHPT-C11}
  E_{\text{kin}}[n]+E_{\text{pair}}[n]
  =\Min_{\rho\leadsto n}
  \Bigl\{\tr\Bigl(\bigl(H_{\text{kin}}+H_{\text{pair}}\bigr)\rho\Bigr)\Bigr\}\,,  
\end{equation}
where
\begin{equation}\label{eq:EHHPT-C12}
  H_{\text{kin}}=\sum_{\sigma}\int(\D\vec{r})\,\psi_{\sigma}(\vec{r})\adj
  \biggl(-\frac{\hbar^2}{2m}\grad^2\biggr)\psi_{\sigma}(\vec{r})
\end{equation}
is the kinetic-energy contribution to $H_{\text{single}}$ in \Eq{C6}. 
Both $E_{\text{kin}}[n]$ and $E_{\text{pair}}[n]$ change when we modify
$V_{\text{int}}(\magn{\vec{r}})$ in $H_{\text{pair}} $.

We incorporate the analog of \Eq{B17} into the functional and have
\begin{align}\label{eq:EHHPT-C13}
  E[n,\mu]&=E[n]+\mu N-\mu\int(\D\vec{r})\,n(\vec{r})\nonumber\\
  &=E_{\text{kin}}[n]
  +\int(\D\vec{r})\,V_{\text{trap}}(\vec{r})n(\vec{r})+E_{\text{pair}}[n]
            \nonumber\\&\pheq\mbox{}
  +\mu N-\mu\int(\D\vec{r})\,n(\vec{r})\,,
\end{align}
whose stationary value is the ground-state energy in accordance with
\Eq{B21}.
The partial derivative $\sds[0pt]{\frac{\partial}{\partial n_a}}$
is the functional derivative
$\sds[0pt]{\frac{\delta}{\delta n(\vec{r})}}$ now, so that
$n_{\text{gs}}(\vec{r})$ and $\mu_{\text{gs}}$ solve
\begin{subequations}
\begin{align}\label{eq:EHHPT-C14a}
  \delta n(\vec{r})
  &:&  \mu-\frac{\delta}{\delta n(\vec{r})}E_{\text{kin}}[n]
  &=V_{\text{trap}}(\vec{r})
  + \frac{\delta}{\delta n(\vec{r})}E_{\text{pair}}[n]\,,\\
  \delta\mu
  &:&  N&=\int(\D\vec{r})\,n(\vec{r})\,,\label{eq:EHHPT-C14b}
\end{align}
\end{subequations}
and $E_{\text{gs}}=E[n_{\text{gs}}]$ follows.

The Hohenberg--Kohn theorem \cite{Hohenberg+Kohn:64} states that different
trapping forces lead to different ground-state densities and, in this sense,
the $V_{\text{trap}}(\vec{r})$ in \Eq{C6} is a functional of $n_{\text{gs}}$,
and then $\rho_{\text{gs}}$ can be regarded as a functional of $n_{\text{gs}}$.
This aspect of DFT, despite its great historical importance, is not central to
the Levy--Lieb constrained-search approach that we are following.
In particular, the functional $E[n,\mu]$ is well-defined also for densities
$n(\vec{r})$ that do not arise as the ground-state densities of a Hamilton
operator $H_{\text{mp}}=H_{\text{single}}+H_{\text{pair}}$ with the ingredients of
\Eq[Eqs.~]{C6} and \Eq[]{C7}.

Mindful of the lessons of the Hartree--Fock method of approximate
many-particle wave functions \cite{Hartree:28,Fock:30} and the
Kohn--Sham scheme in DFT \cite{Kohn+Sham:65}, we regard the right-hand side in
\Eq{C14a} as an effective single-particle potential energy $V(\vec{r})$,
\begin{equation}\label{eq:EHHPT-C15}
  \frac{\delta}{\delta n(\vec{r})}E_{\text{kin}}[n]=\mu-V(\vec{r})\,.
\end{equation}
The Legendre transformation
\begin{equation}\label{eq:EHHPT-C16}
  E_{\text{kin}}[n]\to
  E_{\text{kin}}[n]-\int(\D\vec{r})\,\bigl(\mu-V(\vec{r})\bigr)n(\vec{r})
  =E_1[V-\mu]
\end{equation}
introduces the single-particle functional
\begin{equation}\label{eq:EHHPT-C17}
  E_1[V-\mu]=\tr\Biggl(\biggl(H_{\text{kin}}+\sum_{\sigma}\int(\D\vec{r})\,
  \psi_{\sigma}(\vec{r})\adj
  \bigl(V(\vec{r})-\mu\bigr)\psi_{\sigma}(\vec{r})\biggr)\rho[n]\Biggr)\,.
\end{equation}
We switch from $E_{\text{kin}}[n]$ in the density functional $E[n,\mu]$ to
$E_1[V-\mu]$ in the \emph{density potential functional}
\begin{align}\label{eq:EHHPT-C18}
  E[n,V,\mu]
  &=E_1[V-\mu]
  -\int(\D\vec{r})\,\bigl(V(\vec{r})-V_{\text{trap}}(\vec{r})\bigr)n(\vec{r})
    \nonumber\\
  &\pheq\mbox{}  +E_{\text{pair}}[n]+\mu N\,,
\end{align}
where $n(\vec{r})$, $V(\vec{r})$, and $\mu$ are independent variables;
see also Ref.~\cite{THEchapter}.
The ground-state energy
\begin{equation}\label{eq:EHHPT-C19}
  E_{\text{gs}}(N)=\Ext_{n\leadsto N,V,\mu}\bigl\{ E[n,V,\mu]\bigr\}
  = E[n_{\text{gs}},V_{\text{gs}},\mu_{\text{gs}}]
\end{equation}
is known once $n_{\text{gs}}$, $V_{\text{gs}}$, and $\mu_{\text{gs}}$
are found as the self-consistent solution of
\begin{subequations}\label{eq:EHHPT-C20}
\begin{align}
  \label{eq:EHHPT-C20a}
  \delta n(\vec{r})
  &:& V(\vec{r})
      &=V_{\text{trap}}(\vec{r})
  + \frac{\delta}{\delta n(\vec{r})}E_{\text{pair}}[n]\,,\\
  \label{eq:EHHPT-C20b}
  \delta V(\vec{r})
  &:&n(\vec{r})
      &=\frac{\delta}{\delta V(\vec{r})}E_1[V-\mu]\,,\\
  \label{eq:EHHPT-C20c}
  \delta\mu
  &:& N&=-\pder{}{\mu}E_1[V-\mu]\,.
\end{align}
\end{subequations}
Equation~\Eq[]{C20a} ensures that the effective potential energy equals the
right-hand side of \Eq{C14a}, and Eqs.~\Eq[]{C20b} and \Eq[]{C20c} together
enforce the correct integral of the density $n(\vec{r})$, the constraint in
\Eq{C14b}.

We convert $E[n,V,\mu]$ into $E[n,\mu]$ by enforcing \Eq{C20b}, and get a new
functional $E[V,\mu]$ upon eliminating $n(\vec{r})$ by enforcing \Eq{C20a}.
It is, however, often not possible to perform the conversions to $E[n,\mu]$ or
$E[V,\mu]$ for an actual, approximate, explicit functional $E[n,V,\mu]$.

The main advantage of $E_1[V-\mu]$ over $E_{\text{kin}}[n]$ is that it is
easier to approximate $E_1[V-\mu]$ than $E_{\text{kin}}[n]$.
We note first that
\begin{align}\label{eq:EHHPT-C21}
  E_1[V-\mu]
  &=\sum_{\sigma,\sigma'}\int(\D\vec{r})(\D\vec{r}')\,
  \bra{\vec{r},\sigma}\biggl(\frac{1}{2m}\vec{P}^2+V(\vec{R})-\mu\biggr)
  \ket{\vec{r}',\sigma'}
    \nonumber\\[-1ex]
  &\rule{40pt}{0pt}\mbox{}\times
    \tr\Bigl(\psi_{\sigma'}(\vec{r}')\rho[n]\psi_{\sigma}(\vec{r})\adj\Bigr)
    \nonumber\\
  &=\tr\Bigl((H-\mu)n^{(1)}\Bigr)\,,
\end{align}
where
\begin{equation}\label{eq:EHHPT-C22}
  H=\frac{1}{2m}\vec{P}^2+V(\vec{R})
\end{equation}
is the effective single-particle Hamilton operator and $n^{(1)}$ is the
reduced single-particle statistical operator
\begin{equation}\label{eq:EHHPT-C23}
  n^{(1)}=\sum_{\sigma,\sigma'}\int(\D\vec{r})(\D\vec{r}')\,
  \ket{\vec{r}',\sigma'}
  \,\tr\Bigl(\psi_{\sigma'}(\vec{r}')\rho[n]\psi_{\sigma}(\vec{r})\adj\Bigr)
  \bra{\vec{r},\sigma}\,,
\end{equation}
the version of \Eq{B99a} that applies in the present context.
The traces in \Eq{C23} and the second line in \Eq{C21} are many-particle
traces, whereas the trace in the final line in \Eq{C21} is a single-particle
trace. 

It follows that $E_1[V-\mu]$ is an expectation value of $H-\mu$, a weighted
sum of the eigenvalues of $H-\mu$, and therefore it is the trace of an
operator-valued function of $H-\mu$,
\begin{equation}\label{eq:EHHPT-C24}
  E_1[V-\mu]=\tr\bigl(\mathcal{E}(H-\mu)\bigr)\,.
\end{equation}
The implicit dependence of $E_{\text{kin}}[n]$, and as a consequence also of
${E_1[V-\mu]}$, on $V_{\text{int}}(\magn{\vec{r}})$ --- recall the remark after
\Eq{C12} --- prevents us from stating the actual $\mathcal{E}(H-\mu)$.
For a system of noninteracting fermions, Pauli's exclusion principle implies
that
\begin{equation}\label{eq:EHHPT-C25a}
  \mathcal{E}(H-\mu)=(H-\mu)\eta(\mu-H)\quad\text{when}\
  V_{\text{int}}(\magn{\vec{r}}) =0\,,
\end{equation}
where all single-particle states with ${H<\mu}$ are occupied, those with
${H>\mu}$ are unoccupied, and those with ${H=\mu}$ are partially occupied 
as required by \Eq{C20c},
\begin{equation}\label{eq:EHHPT-C25b}
  N=-\pder{}{\mu}\tr\bigl(\mathcal{E}(H-\mu)\bigr)
  =\tr\bigl(\eta(\mu-H)\bigr)\,;
\end{equation}
$\eta(x)$ denotes Heaviside's unit step function with any value between $0$
and~$1$ for ${x=0}$.
Partial occupation for ${H=\mu}$ occurs when $N$ is noninteger; it can also 
happen for integer $N$ if $\mu$ is a degenerate eigenvalue of $H$.

In the Kohn--Sham scheme, we proceed from
\begin{equation}\label{eq:EHHPT-C26}
  E[n]=E^{(0)}_{\text{kin}}[n]
  +\int(\D\vec{r})\,V_{\text{trap}}(\vec{r})n(\vec{r})
  +E^{\text{(c)}}_{\text{pair}}[n]
\end{equation}
with the density functional
\begin{equation}\label{eq:EHHPT-C27}
  E^{(0)}_{\text{kin}}[n]
  =\Min_{\rho\leadsto n}\bigl\{\tr(H_{\text{kin}}\rho)\bigr\}
\end{equation}
for the kinetic energy of noninteracting
fermions and
\begin{equation}\label{eq:EHHPT-C28}
  E^{\text{(c)}}_{\text{pair}}[n]=E_{\text{pair}}[n]
  +E_{\text{kin}}[n]-E^{(0)}_{\text{kin}}[n]\,,
\end{equation}
which is an amended pair-energy functional that includes the difference
between $E_{\text{kin}}[n]$ and $E^{(0)}_{\text{kin}}[n]$.
If we then use $E^{(0)}_{\text{kin}}[n]$ in \Eq[Eqs.~]{C15} and \Eq[]{C16}
instead of $E_{\text{kin}}[n]$, we arrive at
\begin{align}\label{eq:EHHPT-C29}
  E^{(0)}[n,V,\mu]
  &=E_1^{(0)}[V-\mu]
  -\int(\D\vec{r})\,\bigl(V(\vec{r})-V_{\text{trap}}(\vec{r})\bigr)n(\vec{r})
    \nonumber\\
  &\pheq\mbox{}+E_{\text{pair}}^{\text{(c)}}[n]+\mu N\,,
\end{align}
with
\begin{equation}\label{eq:EHHPT-C30}
  E_1^{(0)}[V-\mu]=\tr\bigl\{(H-\mu)\eta(\mu-H)\bigr\}\,.
\end{equation}
While we now have an explicit expression for $E_1^{(0)}[V-\mu]$ and are no
longer facing the challenge of determining $\mathcal{E}(H-\mu)$, we have the
new task of finding a good approximation for
${E_{\text{kin}}[n]-E^{(0)}_{\text{kin}}[n]}$ in addition to the need of
approximating $E_{\text{pair}}[n]$.
It is common to write
\begin{equation}\label{eq:EHHPT-C31}
  E^{\text{(c)}}_{\text{pair}}[n]=\frac{1}{2}\int(\D\vec{r})(\D\vec{r}')\,
  n(\vec{r})V_{\text{int}}\bigl(\magn{\vec{r}-\vec{r}'}\bigr)n(\vec{r}')
  +E_{\text{xc}}[n]\,,
\end{equation}
the sum of the \emph{Hartree energy} and the \emph{exchange-correlation}
functional $E_{\text{xc}}[n]$.
Many approximations for $E_{\text{xc}}[n]$ have been proposed, some much more
popular than others;
we shall not dive into this ocean and refer the reader to the discussion in
Ref.~\cite{OBchapter}.

The defining feature of the Kohn--Sham scheme is the evaluation of the traces
in \Eq[Eqs.~]{C30}, \Eq[]{C25b}, and 
\begin{align}\label{eq:EHHPT-C32}
  n(\vec{r})
  =\frac{\delta}{\delta V(\vec{r})} E_1^{(0)}[V-\mu]
  &=  \tr\bigl(\delta(\vec{R}-\vec{r})\eta(\mu-H)\bigr)\nonumber\\
  &=\sum_{\sigma}\bra{\vec{r},\sigma}\eta(\mu-H)\ket{\vec{r},\sigma}
\end{align}
in terms of the single-particle eigenstates (``orbitals'') and eigenvalues of
$H$.
In marked contrast, orbital-free DFT aims at approximate expressions for
$E_1[V-\mu]$ or $E_1^{(0)}[V-\mu]$ in conjunction with corresponding
approximations for $E_{\text{pair}}[n]$ or $E^{\text{(c)}}_{\text{pair}}[n]$
in terms of their variables.
Both approaches have their merits.

\section{Momentum-space functionals}\label{sec:EHHPT-4}
One version of nontraditional DFT that has been developed to some extent
\cite{Henderson:81,Buchwald+Englert:89,Englert:92,Cinal+Englert:92,%
  Cinal+Englert:93,vConta+Siedentop:15}
deals with functionals of the momentum-space density
$n(\vec{p})$, where we have 
\begin{equation}
  \psi(a)\to\psi_{\sigma}(\vec{p})\,,
\end{equation}
which annihilates a spin-$\frac12$ fermion with spin label $\sigma$ and
momentum $\vec{p}$.
Then
\begin{subequations}
\begin{align}\label{eq:EHHPT-D2a}
    &\psi_{\sigma}(\vec{p})\psi_{\sigma'}(\vec{p}')\adj
  +\psi_{\sigma'}(\vec{p}')\adj\psi_{\sigma}(\vec{p})
  =\delta_{\sigma,\sigma'}\delta(\vec{p}-\vec{p}')\,,\\ \label{eq:EHHPT-D2b}
  &\mathcal{N}=\sum_{\sigma}\int(\D\vec{p})\,
  \psi_{\sigma}(\vec{p})\adj\psi_{\sigma}(\vec{p})\,,\\ \label{eq:EHHPT-D2c}
  &n(\vec{p})=\tr\biggl(\sum_{\sigma}
  \psi_{\sigma}(\vec{p})\adj\psi_{\sigma}(\vec{p})\rho\biggr)\,,
\end{align}
\end{subequations}
are the analogs of Eqs.~\Eq[]{C3}, \Eq[]{C2}, and \Eq[]{C8}, respectively.
With the Fourier transformed potential energy functions
\begin{equation}\label{eq:EHHPT-D3}
  U_{\text{trap}}(\vec{p})
  =\int\frac{(\D\vec{r})}{(2\pi\hbar)^3}\,\Exp{-\I\vec{p}\cdot\vec{r}/\hbar}
   \,V_{\text{trap}}(\vec{r})
\end{equation}
and
\begin{equation}\label{eq:EHHPT-D4}
  U_{\text{int}}(\magn{\vec{p}})
  =\int\frac{(\D\vec{r})}{(2\pi\hbar)^3}\,\Exp{-\I\vec{p}\cdot\vec{r}/\hbar}
   \,V_{\text{int}}(\magn{\vec{r}})\,,
\end{equation}
the momentum-space versions of \Eq[Eqs.~]{C6} and \Eq[]{C7} read
\begin{align}\label{eq:EHHPT-D5}
  H_{\text{single}}
  &=\sum_{\sigma}\int(\D\vec{p})\,\psi_{\sigma}(\vec{p})\adj
    \frac{1}{2m}\vec{p}^2\psi_{\sigma}(\vec{p})\nonumber\\
  &\pheq\mbox{}
    +\sum_{\sigma}\int(\D\vec{p})(\D\vec{p}')\,
    \psi_{\sigma}(\vec{p})\adj
    U_{\text{trap}}\bigl(\vec{p}-\vec{p}'\bigr)\psi_{\sigma}(\vec{p}')
    \nonumber\\
  &=H_{\text{kin}}+H_{\text{trap}}
\end{align}
and
\begin{equation}\label{eq:EHHPT-D6}
  H_{\text{pair}}=\frac12\sum_{\sigma,\sigma'}
  \int(\D\vec{p})(\D\vec{p}')\,
  \psi_{\sigma}(\vec{p})\adj\psi_{\sigma'}(\vec{p}')\adj
   U_{\text{int}}\bigl(\magn{\vec{p}-\vec{p}'}\bigr)
  \psi_{\sigma'}(\vec{p}')\psi_{\sigma}(\vec{p})  \,.  
\end{equation}
The energy functional of the momentum-space density,
\begin{equation}\label{eq:EHHPT-D7}
  E[n]=\int(\D\vec{p})\,\frac{1}{2m}\vec{p}^2n(\vec{p})
  +E_{\text{trap}}[n]+E_{\text{pair}}[n],
\end{equation}
has an exactly known functional for the kinetic energy
and functionals for the
potential energy of the trapping forces and the pair energy that are jointly
defined by
\begin{align}\label{eq:EHHPT-D8}
  E_{\text{trap}}[n]+E_{\text{pair}}[n]
  &=\Min_{\rho\leadsto n}
    \Bigl\{\tr\bigl((H_{\text{trap}}+H_{\text{pair}})\rho\bigr)\Bigr\}
    \nonumber\\
  &=\tr\bigl(H_{\text{trap}}\rho[n]\bigr)+\tr\bigl(H_{\text{pair}}\rho[n]\bigr)\,,
\end{align}
where ${\rho\leadsto n}$ stands for the constraint in \Eq{D2c}.
Both $E_{\text{trap}}[n]$ and $E_{\text{pair}}[n]$ change when we modify 
$V_{\text{trap}}(\vec{r})$ or $V_{\text{int}}(\magn{\vec{r}})$.%
\footnote{Note that the momentum-space formalism is universal in the dispersal
  relation: We may substitute the quadratic dispersion
  $\vec{p}^2/(2m)$ with, for example, one proportional to
  $\magn{\vec{p}}$, with no change in $E_{\text{trap}}[n]$ and
  $E_{\text{pair}}[n]$.} 

The momentum-space analog of \Eq{C15} is
\begin{equation}
  \frac{\delta}{\delta n(\vec{p})}E_{\text{trap}}[n]=\mu-T(\vec{p})\,,
\end{equation}
which introduces the effective kinetic energy $T(\vec{p})$;
the analogs of \Eq[Eqs.~]{C16} and \Eq[]{C17} are
\begin{align}
  E_1[T-\mu]
  &=E_{\text{trap}}[n]
  -\int(\D\vec{p})\,\bigl(\mu-T(\vec{p})\bigr)n(\vec{p})
  \nonumber\\
  &=\tr\Biggl(\biggl(H_{\text{trap}}+\sum_{\sigma}
  \int(\D\vec{p})\,\psi_{\sigma}(\vec{p})\adj\bigl(T(\vec{p})-\mu\bigr)
  \psi_{\sigma}(\vec{p})\biggr)\rho[n]\Biggr)\,,
\end{align}
which takes us to the analog of \Eq{C18},
\begin{align}
  E[n,T,\mu]&=E_1[T-\mu]-\int(\D\vec{p})\,
              \biggl(T(\vec{p})-\frac{1}{2m}\vec{p}^2\biggr)n(\vec{p})
              \nonumber\\
  &\pheq\mbox{}+E_{\text{pair}}[n]+\mu N\,.
\end{align}
The ground-state values $n_{\text{gs}}$, $T_{\text{gs}}$, $\mu_{\text{gs}}$
are the self-consistent solution of
\begin{align}\label{eq:EHHPT-D12}
  \delta n(\vec{p})
  &:& T(\vec{p})
  &=\frac{1}{2m}\vec{p}^2
    +\frac{\delta}{\delta n(\vec{p})}E_{\text{pair}}[n]\,,
  \nonumber\\                
  \delta T(\vec{p})
  &:& n(\vec{p})
  &=\frac{\delta}{\delta T(\vec{p})}E_1[T-\mu]\,,\nonumber\\
  \delta \mu&:& N&=-\pder{}{\mu}E_1[T-\mu]\,,
\end{align}
which are the momentum-space analogs of \Eq[Eqs.~]{C20}.

Further, the analog of \Eq{C24} reads
\begin{equation}\label{eq:EHHPT-D13}
  E_1[T-\mu]=\tr\bigl(\mathcal{E}(H-\mu)\bigr)
\end{equation}
with
\begin{equation}\label{eq:EHHPT-D14}
  H=T(\vec{P})+V_{\text{trap}}(\vec{R})\,,
\end{equation}
and \Eq{C25a} continues to apply for noninteracting fermions.
The momen\-tum-space Kohn--Sham scheme is fully analogous to that of
\Eq[Eqs.~]{C26}--\Eq[]{C32} in configuration space;
details can be found in Ref.~\cite{Cinal+Englert:93}.

We note that the single-particle Hamilton operators in \Eq[Eqs.~]{C22} and
\Eq[]{D14} do not commute and, therefore, have different sets of Kohn--Sham
orbitals.
It follows that the Kohn--Sham orbitals cannot be L\"owdin's ``natural
orbitals'' \cite{Lowdin:55}, the eigenstates of $n^{(1)}$ for
${\rho[n]=\rho[n_{\text{gs}}]}$ in \Eq{C23}.

\section{\protect\mbox{%
    Thomas--Fermi atoms in configuration and momentum space}}
\label{sec:EHHPT-5}
When the spin-$\frac12$ fermions are electrons in an atom with nuclear charge
$Ze$, we have a Coulomb potential for both $V_{\text{trap}}(\vec{r})$ and
$V_{\text{int}}(\magn{\vec{r}})$,
\begin{equation}
   V_{\text{trap}}(\vec{r})=-\frac{Ze^2}{\magn{\vec{r}}}\,,\qquad
   V_{\text{int}}(\magn{\vec{r}})=\frac{e^2}{\magn{\vec{r}}}\,,
\end{equation}
for which
\begin{equation}
  U_{\text{trap}}(\vec{p})=-\frac{Ze^2}{2\pi^2\hbar}\frac{1}{\vec{p}^2}\,,
  \qquad
   U_{\text{int}}(\magn{\vec{p}})=\frac{e^2}{2\pi^2\hbar}\frac{1}{\vec{p}^2}
\end{equation}
are the respective Fourier transforms.
In the Thomas--Fermi model \cite{Thomas:27,Fermi:28} for atoms we use the
leading semiclassical approximations for $E_1[V-\mu]$ and $E_1[T-\mu]$, with
${\mu<0}$ here, and approximate $E_{\text{pair}}[n]$ by the dominant Hartree
energy. 
The latter is
\begin{equation}
  E^{(\text{TF})}_{\text{pair}}[n]
  =\frac12\int(\D\vec{r})(\D\vec{r}')\,
  n(\vec{r})\frac{e^2}{\magn{\vec{r}-\vec{r}'}}n(\vec{r}')
\end{equation}
in configuration space and
\begin{equation}
    E^{(\text{TF})}_{\text{pair}}[n]
    =\frac12\int(\D\vec{p})(\D\vec{p}')\,
    \bigl(3\pi^2\bigr)^{2/3} \frac{e^2}{2\pi^2\hbar}
    \biggl(n_>^{2/3}n_<^{\,}-\frac{1}{5}n_<^{5/3}\biggr)
\end{equation}
with ${n_>=\Max\bigl\{n(\vec{p}),n(\vec{p}')\bigr\}}$ and
${n_<=\Min\bigl\{n(\vec{p}),n(\vec{p}')\bigr\}}$ in momentum space.

The semiclassical approximation of the single-particle functionals replaces
the quantum mechanical traces in \Eq[Eqs.~]{C24} and \Eq[]{D13} by classical
phase space integrals after approximating $E_1[V-\mu]$ and $E_1[T-\mu]$ by the
respective Kohn--Sham expressions.
Accordingly, we have
\begin{equation}\label{eq:EHHPT-E5}
  E_1^{(\text{TF})}[V-\mu]=2\int\frac{(\D\vec{r})(\D\vec{p})}{(2\pi\hbar)^3}
  \biggl(\frac{\vec{p}^2}{2m}+V(\vec{r})-\mu\biggr)
    \eta\biggl(\mu-\frac{\vec{p}^2}{2m}-V(\vec{r})\biggr)
\end{equation}
in configuration space and
\begin{equation}\label{eq:EHHPT-E6}
  E_1^{(\text{TF})}[T-\mu]=2\int\frac{(\D\vec{r})(\D\vec{p})}{(2\pi\hbar)^3}
  \biggl(T(\vec{p})-\frac{Ze^2}{\magn{\vec{r}}}-\mu\biggr)
  \eta\biggl(\mu-T(\vec{p})+\frac{Ze^2}{\magn{\vec{r}}}\biggr)
\end{equation}
in momentum space, where the factor of two is the spin multiplicity.
Upon evaluating the $\vec{p}$ integral in \Eq{E5} and the $\vec{r}$ integral
in \Eq{E6} we arrive at the respective Thomas--Fermi functionals,
\begin{align}\label{eq:EHHPT-E7}
  E^{(\text{TF})}[n,V,\mu]
  &=-\int(\D\vec{r})\,\frac{1}{15\pi^2\hbar^3m}
    \Bigl[2m\bigl(\mu-V(\vec{r})\bigr)\Bigr]_+^{5/2}\nonumber\\
  &\pheq\mbox{}
    -\int(\D\vec{r})\,\biggl(V(\vec{r})
    +\frac{Ze^2}{\magn{\vec{r}}}\biggr)n(\vec{r})
    \nonumber\\
  &\pheq\mbox{}
    +\frac12\int(\D\vec{r})(\D\vec{r}')\,
  n(\vec{r})\frac{e^2}{\magn{\vec{r}-\vec{r}'}}n(\vec{r}')+\mu N
\end{align}
with ${[x]_+=x\eta(x)}$ and
\begin{align}\label{eq:EHHPT-E8}
  E^{(\text{TF})}[n,T,\mu]
  &=-\int(\D\vec{p})\,\frac{1}{6\pi^2}
  \frac{(Ze^2/\hbar)^3}{\bigl(T(\vec{p})-\mu\bigr)^2}  \nonumber\\
  &\pheq\mbox{}-\int(\D\vec{p})\,
  \biggl(T(\vec{p})-\frac{\vec{p}^2}{2m}\biggr)n(\vec{p}) \nonumber\\
  &\pheq\mbox{}
  +\frac12\int(\D\vec{p})(\D\vec{p}')\,
    \biggl(\frac3\pi\biggr)^{2/3} \frac{e^2}{2\hbar}
    \biggl(n_>^{2/3}n_<^{\,}-\frac{1}{5}n_<^{5/3}\biggr)+\mu N\,.
\end{align}
Although these functionals are obviously quite different in structure, they
are equivalent in their implications because the two step functions in
\Eq[Eqs.~]{E5} and \Eq[]{E6} select the same classically allowed region in
phase space for the stationary values of $V(\vec{r})$, $T(\vec{p})$, and $\mu$,
\begin{equation}
  \eta\biggl(\mu_{\text{gs}}-\frac{\vec{p}^2}{2m}
             -V_{\text{gs}}(\vec{r})\biggr)
  =\eta\biggl(\mu_{\text{gs}}-T_{\text{gs}}(\vec{p})
              +\frac{Ze^2}{\magn{\vec{r}}}\biggr)\,;
\end{equation}
see Refs.~\cite{Englert:92,vConta+Siedentop:15} for the details.

Once we improve on the Thomas--Fermi approximation and include the leading
correction --- the Scott correction \cite{Scott:52} for the strongly bound
electrons, that is --- the respective Thomas--Fermi--Scott models
\cite{Englert+Schwinger:84a,Cinal+Englert:92} are not equivalent.
The same remark applies at the next level of approximation where we account
for the exchange energy and the leading quantum correction to the phase space
integrals in \Eq[Eqs.~]{E5} and \Eq[]{E6}.

\section{Single-particle-exact  functionals}\label{sec:EHHPT-6}
The configuration-space functional $E_{\text{single}}[n]$ in \Eq{C10} treats
the potential energy of the trapping forces exactly and, in all practical
applications, requires a good approximation for the kinetic-energy
contribution $E_{\text{kin}}[n]$.
The momentum-space functional in \Eq{D7} is exact for the kinetic energy but
needs an approximation for $E_{\text{trap}}[n]$.
It is also possible to have an exact density functional for both terms in
${E_{\text{single}}[n]=E_{\text{kin}}[n]+E_{\text{trap}}[n]}$, not just for
one or the other.

For this purpose, we choose modes that refer to the eigenstates of
$H_{\text{single}}$, so that
\begin{equation}
  \ket{a}\to\ket{k,\sigma}\,,\qquad\psi(a)\to\psi_{k,\sigma}\,,
\end{equation}
where
\begin{equation}
  H_{\text{1p}}(\vec{R},\vec{P})\ket{k,\sigma}=\ket{k,\sigma}\varepsilon_k
  \quad\text{for}\quad k=0,1,2,\dots
\end{equation}
with ${\varepsilon_k\geq\varepsilon_{k'}}$ when ${k>k'}$ for the
spin-degenerate eigenvalues.
Then, the occupation numbers
\begin{equation}\label{eq:EHHPT-F3}
  n_k=\sum_{\sigma}
  \tr\bigl(\psi_{k,\sigma}\adj\psi_{k,\sigma}\phadj\rho\bigr)
\end{equation}
are restricted by ${0\leq n_k\leq2}$ for all $k$.%
\footnote{There could be scattering states with continuous eigenvalues in
  addition to the discrete bound states of $H_{\text{1p}}$ that we are
  considering.
  The completeness relation in \Eq{B3'} needs the scattering states as well as
  the bound states.
  Following the practice of Kohn--Sham calculations, we regard the scattering
  states as unoccupied.}
The energy functional of \Eq{B19} is
\begin{equation}
  E[n]=\sum_{k=0}^{\infty}\varepsilon_kn_k+E_{\text{pair}}[n]
\end{equation}
with
\begin{equation}\label{eq:EHHPT-F5}
  E_{\text{pair}}[n]
  =\Min_{\rho\leadsto n}\bigl\{\tr(H_{\text{pair}}\rho)\bigr\}
  =\tr\bigl(H_{\text{pair}}\rho[n]\bigr)\,,
\end{equation}
where ${\rho\leadsto n}$ enforces the constraint of \Eq{F3}.
Indeed, we have an exact density functional for $E_{\text{single}}[n]$ and, as
always, need to approximate $E_{\text{pair}}[n]$.

Very much of this territory is unexplored.
Work on approximate pair-energy functionals is ongoing and progressing
\cite{Huang:22,Trappe+alii:22}.
We have, for example, an algorithm for generating a full single-particle
density matrix from the prechosen diagonal elements, the occupation numbers in
\Eq{F3}, as well as a Thomas--Fermi-type approximation for the off-diagonal
matrix elements.
With that at hand, we follow Dirac's guidance \cite{Dirac:30} and
arrive at a two-particle density matrix in Hartree--Fock approximation, which
in turn yields a value for the trace in \Eq{F5}.
Alternatively, when the pair interaction is weak, we can approximate
$E_{\text{pair}}[n]$ in second-order perturbation theory \cite{JCprivate}.

\setcounter{equation}{0}\renewcommand{\theequation}{A.\arabic{equation}}
\setcounter{theorem}{0}\renewcommand{\thetheorem}{A.\arabic{theorem}}
  
\section*{Appendix: Semiclassical eigenvalues}
\addcontentsline{toc}{section}{Appendix: Semiclassical eigenvalues}
Semiclassical approximations --- and their unreasonable accuracy --- are
central to DFT as emphasized by Okun and Burke in Ref.~\cite{OBchapter}.
The phase space integrals in \Eq[Eqs.~]{E5} and \Eq[]{E6} are to the point.

Approximate eigenvalues of Hamilton operators, obtained by the WKB method
\cite{Wentzel:26,Kramers:26,Brillouin:26}, play an important role, too.
For one-dimensional Hamilton operators of the standard form
\begin{equation}\label{eq:EHHPT-Z1}
  H(X,P)=\frac{1}{2m}P^2+V(X)
\end{equation}
we find an approximation for the $k$th eigenvalue by the well-known
quantization rule
\begin{align}\label{eq:EHHPT-Z2}
  k+\frac12
  &=\int\frac{\D x\,\D p}{2\pi\hbar}\,\eta\bigl(E_k-H(x,p)\bigr)\nonumber\\
  &=\frac{1}{\pi\hbar}\int\D x\,\Bigl[2m\bigl(E_k-V(x)\bigr)\Bigr]^{1/2}_+
\end{align}
with ${k=0,1,2,\dots}$~.
The corresponding expression for an eigenvalue of the isotropic
three-dimensional Hamilton operator
\begin{equation}\label{eq:EHHPT-Z3}
  H(\vec{R},\vec{P})=\frac{1}{2m}\vec{P}^2+V\bigl(\magn{\vec{R}}\bigr)
\end{equation}
is equally familiar,
\begin{equation}\label{eq:EHHPT-Z4}
  k+\frac12=\frac{1}{\pi\hbar}\int\D r\,\biggl[2m\biggl(E_{k,l}
  -\frac{\hbar^2}{2m}\frac{(l+\frac12)^2}{r^2}-V(r)\biggl)\biggr]_+^{1/2}\,,
\end{equation}
where ${k=0,1,2,\dots}$ is the radial quantum number and ${l=0,1,2,\dots}$ is
the angular momentum quantum number, as the sector with
$\vec{L}^2=(\vec{R}\boldsymbol{\times}\vec{P})^2=l(l+1)\hbar^2$ is considered. 
The replacement ${l(l+1)\to(l+\frac12)^2}$ is the so-called Langer correction
\cite{Langer:37}.
It is remarkable that \Eq{Z4} yields the exact Bohr energies for the Coulomb
potential ${V(r)=-Ze^2/r}$, rather than having systematically small errors in
the correspondence limit of large quantum numbers.
The performance of \Eq{Z4} is better than one should reasonably expect.

Let us put spin multiplicity aside and consider
\begin{equation}\label{eq:EHHPT-Z5}
  \nu(\varepsilon)
  =\tr\Bigl(\eta\bigl(\varepsilon-H(\vec{R},\vec{P})\bigr)\Bigr)\,,
\end{equation}
which is the count of eigenvalues of $ H(\vec{R},\vec{P})$ below the threshold
$\varepsilon$.
The value of $\nu(\varepsilon)$ is an integer when $\varepsilon$ is between
successive eigenvalues and equals any intermediate noninteger when
$\varepsilon$ is an eigenvalue of $H(\vec{R},\vec{P})$;
the graph of $\nu(\varepsilon)$ is a stair case.
The evaluation of the trace by a phase space integral,
\begin{equation}\label{eq:EHHPT-Z6}
  \nu(\varepsilon)=\int\frac{(\D\vec{r})(\D\vec{p})}{(2\pi\hbar)^3}\,
  \Bigl[\eta\bigl(\varepsilon-H(\vec{R},\vec{P})\bigr)
  \Bigr]_{\textsc{w}}(\vec{r},\vec{p})\,,
\end{equation}
involves the Wigner function \cite{Wigner:32,Englert:89}, sometimes called
Weyl symbol, of the operator 
$\eta(\varepsilon-H)$.
For an operator-valued function of $H$ we have the semiclassical approximation
\begin{equation}\label{eq:EHHPT-Z7}
  \bigl[f(H)\bigr]_{\textsc{w}}(\vec{r},\vec{p})
  \cong f\bigl(H_{\textsc{w}}(\vec{r},\vec{p})\bigr)\,,
\end{equation}
where we neglect all quantum corrections --- technically speaking, these are
terms involving even powers of Planck's constant times the Poisson-bracket
differential operator; see, for example, the appendix in
Ref.~\cite{vonEiff+Weigel:91}.  

Quite generally, then,
\begin{align}\label{eq:EHHPT-Z8}
  \tr\Bigl(f\bigl(H(\vec{R},\vec{P})\bigr)\Bigr)
  &=\int\frac{(\D\vec{r})(\D\vec{p})}{(2\pi\hbar)^3}\,
  \Bigl[f\bigl(H(\vec{R},\vec{P})\bigr)\Bigr]_{\textsc{w}}(\vec{r},\vec{p})
  \nonumber\\
  &\cong\int\frac{(\D\vec{r})(\D\vec{p})}{(2\pi\hbar)^3}\,
  f\bigl(H_{\textsc{w}}(\vec{r},\vec{p})\bigr)
\end{align}
provides a semiclassical approximation for the trace of a function of the
Hamilton operator, and this can be improved by including some of the quantum
corrections.
We have examples of \Eq{Z8} in \Eq[Eqs.~]{E5} and \Eq[]{E6} since
\begin{equation}
  \bigl[T(\vec{P})+V(\vec{R})\bigr]_{\textsc{w}}(\vec{r},\vec{p})
  =T(\vec{p})+V(\vec{r})\,.
\end{equation}

When we apply \Eq{Z7} to \Eq{Z6}, we approximate the stair-case function by a
very smooth function of $\varepsilon$, which suggests to determine approximate
eigenvalues of $H$ by 
\begin{equation}\label{eq:EHHPT-Z10}
  k+\frac12=\int\frac{(\D\vec{r})(\D\vec{p})}{(2\pi\hbar)^3}\,
  \eta\bigl(E_k-H_{\textsc{w}}(\vec{r},\vec{p})\bigr)\,.
\end{equation}
This works best for the one-dimensional Hamilton operators of \Eq{Z1}
for which \Eq{Z10} becomes \Eq{Z2}.

The application to an angular-momentum sector of an isotropic
three-dimensional Hamilton operator as in \Eq{Z3} requires a proper reduction
to an effective one-dimensional Hamilton operator for the radial motion.
With
\begin{equation}\label{eq:EHHPT-Z11}
  R=\magn{\vec{R}}\,,\qquad
  \Gamma=\frac12\bigl(\vec{R}\cdot\vec{P}+\vec{P}\cdot\vec{R}\bigr)\,,\qquad
  [R,\Gamma]=\I\hbar R
\end{equation}
we have
\begin{equation}
  \vec{P}^2
  =\frac1R\Bigl(\Gamma^2+\vec{L}^2+\tfrac{1}{4}\hbar^2\Bigr)\frac1R
  \to\frac1R\Bigl(\Gamma^2+\bigl(l+\tfrac12\bigr)^2\hbar^2\Bigr)\frac1R\,,
\end{equation}
where we restrict to an angular-momentum sector in the last step, and
recognize the origin of the Langer correction.
For the Hamilton operator in \Eq{Z3} this means
\begin{equation}
  H(\vec{R},\vec{P})\to H_l(R,\Gamma)
  =\frac{1}{2m}\frac1R\Bigl(\Gamma^2+\bigl(l+\tfrac12\bigr)^2\hbar^2\Bigr)\frac1R
  +V(R)\,.
\end{equation}
We introduce a proper Heisenberg $X,P$ pair in accordance with
\begin{equation}
  \kappa R=\Exp{\kappa X}\,,\qquad \kappa\Gamma=P\,,\qquad
  [X,P]=\I\hbar\,,
\end{equation}
where $\kappa$ is a reference wave number, a reciprocal length, that ensures
the correct metrical dimensions, and obtain
\begin{equation}
  H_l^{\text{(1D)}}(X,P)=\Exp{-\kappa X}\frac{P^2}{2m}\Exp{-\kappa X}
  +\frac{(\hbar\kappa)^2}{2m}\bigl(l+\tfrac12\bigr)^2\Exp{-2\kappa X}
  +V\Bigl(\kappa^{-1}\Exp{\kappa X}\Bigr)\,.
\end{equation}
The replacements $X\to x$, $P\to p$ turn this $H_l^{\text{(1D)}}(X,P)$ into
its Wigner function, so that
\begin{equation}
  k+\frac12=\int\frac{\D x\,\D p}{2\pi\hbar}\,
  \eta\Bigl(E_{k,l}-H_l^{\text{(1D)}}(x,p)\Bigr)
\end{equation}
is the proper analog of \Eq{Z10}.
After evaluating the $p$ integral and switching from $x$ to
${r=\kappa^{-1}\Exp{\kappa x}}$, we arrive at \Eq{Z4}.

This procedure does not treat $\vec{R}$ and $\vec{P}$ on equal footing.
Rather than giving a privileged role to the position operator in \Eq{Z11} we
can just as well single out the momentum operator and work with the pair
${P=\magn{\vec{P}}}$ and $\Gamma$.
For ${V(r)=-Ze^2/r}$ this ``semiclassical quantization in momentum space''
does not yield the exact Bohr energies.
Instead, we get very good approximate eigenvalues with systematically smaller
errors for larger quantum numbers; see Ref.~\cite{Rohwedder+Englert:94} for
the details.

\section*{Acknowledgments}\addcontentsline{toc}{section}{Acknowledgments}
We have benefitted profoundly from discussions with
Kieron Burke,
Jerzy Cios\l{}owski,
John Dobson,
Reiner Dreizler,
Yuan Ping Feng,
Mel Levy,
Heinz Siedentop,
Giovanni Vignale,
Su Ying Quek, and
Weitao Yang.
We thank them sincerely.


\begin{thebibliography}{99}

\bibitem{HSchapter}
H. Siedentop,
``Mathematical elements of density functional theory'',
Chapter~1, pp.~1--55 in \cite{DFMPS-2019};
arXiv:2203.14069 [math-ph].

\bibitem{OBchapter}
P.~Okun and K.~Burke,
``Semiclassics: The hidden theory behind the success of DFT'',
Chapter~7, pp.~179--248  in \cite{DFMPS-2019};
arXiv:2105.04384 [physics.chem-ph].

\bibitem{Compton:23}
A. H. Compton, ``The Spectrum of Scattered X-Rays'',
\textit{Phys.\ Rev.\/} \textbf{22}, 409--413 (1923).  

\bibitem{Henderson:81}
G. A. Henderson,
``Variational theorems for the single-particle probability density and density
matrix in momentum space'', 
\textit{Phys.\ Rev.\ A} \textbf{23}, 19--20 (1981).

\bibitem{Buchwald+Englert:89}
K. Buchwald and B.-G. Englert,
``Thomas--Fermi--Scott model: Momentum-space density'',
\textit{Phys.\ Rev.\ A} \textbf{40}, 2738--2741 (1989).

\bibitem{Englert:92}
B.-G. Englert,
``Energy functionals and the Thomas--Fermi model in momentum space'',
\textit{Phys.\ Rev.\ A} \textbf{45}, 127--134 (1992).

\bibitem{Cinal+Englert:92}
M. Cinal and B.-G. Englert,
``Thomas--Fermi--Scott model in momentum space'',
\textit{Phys.\ Rev.\ A} \textbf{45}, 135--139 (1992).

\bibitem{Cinal+Englert:93}
M. Cinal and B.-G. Englert,
``Energy functionals in momentum space: Exchange energy, quantum corrections,
and the Kohn--Sham scheme'',
\textit{Phys.\ Rev.\ A} \textbf{48}, 1893--1902 (1993).

\bibitem{vConta+Siedentop:15}
V. von Conta and H. Siedentop,
``Statistical theory of the atom in momentum space'',
\textit{Markov Process.\ Relat.\ Fields} \textbf{21}, 433--448 (2015).

\bibitem{Levy:79}
M. Levy, ``Universal variational functional of electron densities,
first-order density matrices and solution to the $v$-representability
problem'', 
\textit{Proc.\ Natl.\ Acad.\ Sci.\ U.S.A.\/}
\textbf{76}, 6062--6065 (1979).

\bibitem{Lieb:83}
E. H. Lieb, ``Density functionals for Coulomb systems'',
\textit{Int.\ J. Quantum Chem.\/} \textbf{24}, 243--277 (1983).

\bibitem{Schwinger:QMbook}
J. Schwinger,
\textit{Quantum Mechanics --- Symbolism of Atomic Measurements},
B.-G. Englert (ed.), 2nd corrected printing,
Springer, Heidelberg, 2003.

\bibitem{Goral+2:01}
K. G\'oral, B.-G. Englert, and K. Rz\c{a}\.zewski,
``Semiclassical theory of trapped fermionic dipoles'',
\textit{Phys.\ Rev.\ A} \textbf{63}, 033606 (2001).

\bibitem{Fang+Englert:11}
B. Fang and B.-G. Englert,
``Density functional of a two-dimensional gas of dipolar atoms:
Thomas--Fermi--Dirac treatment'',
\textit{Phys.\ Rev.\ A} \textbf{83}, 052517 (2011).

\bibitem{RMDchapter}
R. M. Dreizler,
``Remarks on the density functional theory of relativistic
many-particle systems'',
Chapter~9, pp.~267--283 in \cite{DFMPS-2019}.

\bibitem{Perdew+3:82}
J. P. Perdew, R. G. Parr, M. Levy, and J. L. Balduz, Jr.,
``Density-functional theory for fractional particle number:
Derivative discontinuities of the energy'',
\textit{Phys.\ Rev.\ Lett.\/} \textbf{49}, 1691--1694 (1982).

\bibitem{Perdew+3:00}
J. P. Perdew, R. G. Parr, M. Levy, and J. L. Balduz, Jr.,
``Perspective on `Density-functional theory for fractional particle number:
Derivative discontinuities of the energy'\,'',
\textit{Theor.\ Chim.\ Acc.\/} \textbf{103}, 346--348 (2000).

\bibitem{Yang+2:00}
W. Yang, Y. Zhang, and P. W. Ayers,
``Degenerate ground states and a fractional number of electrons in density and
reduced density matrix functional theory'',  
\textit{Phys.\ Rev.\ Lett.\/} \textbf{84}, 5172--5175 (2000).

\bibitem{Baerends:20}
E. J. Baerends,
``On derivatives of the energy with respect to total electron number and
orbital occupation numbers. A critique of Janak's theorem'',
\textit{Mol.\ Phys.\/} \textbf{118}, e1612955 (2020).

\bibitem{LZGchapter}
M.~Levy, F.~Zahariev, and M.~S.~Gordon,
``Spin-density functional theory through spin-free wave functions'',
Chapter~11, pp.~307--316 in \cite{DFMPS-2019}.

\bibitem{Hohenberg+Kohn:64}
P.~Hohenberg and W.~Kohn,
``Inhomogeneous electron gas'',
\textit{Phys.\ Rev.\/} \textbf{136}, B864--B871 (1964).  

\bibitem{Hartree:28}
D. Hartree,
``The wave mechanics of an atom with a non-Coulomb central
field. Part II. Some results and discussion'',
\textit{Math.\ Proc.\ Cambridge Phil.\ Soc.} \textbf{24}, 111--132 (1928).

\bibitem{Fock:30}
V.~Fock,
``N\"aherungsmethode zur L\"osung des quantenmechanischen
Mehr\-k\"orper\-problems'',  Z. Physik \textbf{61}, 126--148 (1930).

\bibitem{Kohn+Sham:65}
W.~Kohn and L.~J. Sham,
``Self-consistent equations including exchange and correlation effects'',
\textit{Phys.\ Rev.\/} \textbf{140}, A1133--A1138 (1965).

\bibitem{THEchapter}
M.-I. Trappe, J.~H. Hue, and B.-G. Englert,
``Density-potential functional theory for fermions in one dimension'',
Chapter~8, pp.~249--265 in \cite{DFMPS-2019};
arXiv:2106.07839 [cond-mat.quant-gas].

\bibitem{Lowdin:55}
P.-O. L\"owdin,
``Quantum theory of many-particle Systems. I. Physical interpretations by
means of density matrices, natural spin-orbitals, and convergence problems in
the method of configurational interaction'', 
\textit{Phys.\ Rev.\/} \textbf{97}, 1474--1489 (1955).

\bibitem{Thomas:27}
L.~H. Thomas, ``The calculation of atomic fields'', \textit{Math. Proc.
  Cambridge Philos. Soc.} \textbf{23}, 542--548 (1927).

\bibitem{Fermi:28}
E.~Fermi, ``Eine statistische Methode zur Bestimmung einiger
  Eigenschaften des Atoms und ihre Anwendung auf die Theorie des periodischen
  Systems der Elemente'', \textit{Z. Physik} \textbf{48}, 73--79 (1928).

\bibitem{Scott:52}
J.~M.~C.~Scott,
``The binding energy of the Thomas-Fermi atom'',
\textit{Phil. Mag.} \textbf{43}, 859--867 (1952).

\bibitem{Englert+Schwinger:84a}
B.-G. Englert and J.~Schwinger,
``Statistical atom: Handling the strongly bound electrons'',
\textit{Phys.\ Rev.\ A} \textbf{29}, 2331--2338 (1984).

\bibitem{Huang:22}
Z.~C.~Huang,
\textit{Single Particle Exact Density Functionals},
B.Sc.\ thesis, National University of Singapore, 2022.

\bibitem{Trappe+alii:22}
M.-I. Trappe et al.\ (in preparation).
 
\bibitem{Dirac:30}
P.~A.~M. Dirac,
``Note on exchange phenomena in the Thomas atom'',
\textit{Math. Proc. Cambridge Philos. Soc.} \textbf{26}, 376--385 (1930).
 
\bibitem{JCprivate}
J. Cios\l{}owski, private communication.

\bibitem{Wentzel:26}
G.~Wentzel, ``Eine Verallgemeinerung der Quantenbedingungen f{\"u}r die
  Zwecke der Wellenmechanik'', \textit{Z. Physik} \textbf{38}, 518--529 (1926).

\bibitem{Kramers:26}
H.~A. Kramers, ``Wellenmechanik und halbzahlige Quantisierung'',
  \textit{Z. Physik} \textbf{39}, 828--840 (1926).

\bibitem{Brillouin:26}
  L.~Brillouin, ``La m\'ecanique ondulatoire de Schr\"odinger:
    une m\'ethode generale de resolution par approximations successives'',
  \textit{Compt. Rend.} \textbf{183}, 24 (1926).

\bibitem{Langer:37}
R. E. Langer,
``On the connection formulas and the solutions of the wave equation'',
\textit{Phys.\ Rev.\/} \textbf{51}, 669--676 (1937).

\bibitem{Wigner:32}
E. Wigner,
``On the quantum correction for thermodynamic equilibrium'',
\textit{Phys.\ Rev.\/} \textbf{40}, 749--759 (1932).

\bibitem{Englert:89}
B.-G. Englert,
``On the operator bases underlying Wigner's, Kirkwood's and Glauber's phase
space functions'',
\textit{J. Phys.\ A: Math.\ Gen.\/} \textbf{22}, 625--640 (1989)

\bibitem{vonEiff+Weigel:91}
D. von Eiff and M. K. Weigel, ``The relativistic Hartree--Fock approximation
for finite temperatures and its semiclassical expansion'',
\textit{Z. Physik A} \textbf{339}, 63--70 (1991). 

\bibitem{Rohwedder+Englert:94}
B. Rohwedder and B.-G. Englert,
``Semiclassical quantization in momentum space'',
\textit{Phys.\ Rev.\ A} \textbf{49}, 2340--2346 (1994).

\bibitem{DFMPS-2019}
\textit{Density Functionals for Many-Particle Systems:
Mathematical Theory and Physical Applications of Effective Equations},
B.-G. Englert, H. Siedentop, and M.-I. Trappe (eds.),
Lect.\ Notes Ser., IMS, NUS \textbf{41}, World Scientific, Singapore, 2022.

\end{thebibliography}
\end{document}